\def\der{\partial}
\def\la{\lambda}
\def\bla{\bar{\lambda}}
\def\las{\{ \la_i \}}
\def\mus{\{ \mu_k \}}
\def\A{{\bf A}}
\def\B{{\bf B}}
\def\C{{\bf C}}
\def\D{{\bf D}}
\def\F{{\bf R}}
\def\T{{\bf T}}

\def\i{{\rm i}}
\def\d{{\rm d}}
\def\e{{\rm e}}

\def\bra#1{\left\langle #1 \right|}
\def\ket#1{\left| #1 \right\rangle}
\def\bracket#1#2{\left\langle #1 \right|\left.\! #2 \right\rangle}
\def\Im{\mathop{\rm Im}\nolimits} 
\def\dps{\displaystyle}
\documentstyle[12pt,cite,amssymb,epsfig]{article}
\begin{document}
\title{\large
\bf Six - Vertex Model with Domain wall boundary
conditions. Variable inhomogeneities.} 
\author{J.~de~Gier \\
{\it Department of Mathematics}\\
{\it School of Mathematical Sciences}\\
{\it Australian National University}\\
{\it Canberra ACT 0200, Australia}\\[29pt]
V.~Korepin\\
{\it C.N.~Yang Institute for Theoretical Physics}\\
{\it State University of New York at Stony Brook}\\
{\it Stony Brook, NY 11794--3840, USA}}
\date{\today}
\maketitle 
\begin{abstract}
We consider the six-vertex model with domain wall boundary conditions.
We choose the inhomogeneities as solutions of the Bethe Ansatz equations.
The Bethe Ansatz equations have many solutions, so we can consider a wide
variety of  inhomogeneities. For certain choices of the inhomogeneities we
study arrow correlation functions on the horizontal line going through the 
centre. In particular we obtain a multiple integral representation for the 
emptiness formation probability that generalizes the known formul\ae\ for 
XXZ antiferromagnets.
\end{abstract}

\section{Introduction}
The six-vertex model was first introduced in \cite{Sla}.
It was solved exactly by E. Lieb \cite{Lieb} and B.~Sutherland 
\cite{Suth} in 1967 by means of a Bethe Ansatz for periodic
boundary  conditions. Later the six-vertex model was studied 
also in the presence of several other boundary conditions 
\cite{BBOY,OB,Elo}. Domain wall boundary conditions were introduced 
in 1982 \cite{Kor}. These boundary conditions are interesting because 
they allow for the derivation of determinant representations for 
correlation functions \cite{BIK}. It was realised recently that the 
six-vertex model in the presence of such boundary conditions is extremely
helpful in the enumeration of alternating sign matrices \cite{Kup,Br}.
The bulk free energy for these boundary conditions was calculated in 
\cite{KZJa}.

In this paper we show that for special choices of inhomogeneities,
one can compute the free energy and some correlation functions of the 
system. This observation might be useful because we expect some properties 
of the model to be independent of the inhomogeneities i.e.\ to depend only 
on the anisotropy parameter. In the simplest situation, the correlation 
functions coincide with the ones for periodic boundary conditions. We will 
continue the line of research of \cite{KZJb} and will primarily be interested
in the emptiness formation probability (EFP) which was first introduced in 
\cite{KIEU}. In the latter paper a multiple integral expression for the EFP 
was obtained for the first time for the ground state of the XXX antiferromagnet.
Here we will consider the EFP for more general Bethe states and will generalize
the multiple integral expression to these cases.

\section{Quantum inverse scattering}

\subsection{Setup and notation}
We consider the inhomogeneous six-vertex model with domain wall boundary
conditions. It is defined as the six-vertex model on a $M\times M$
square lattice with fixed boundary conditions: arrows on the
horizontal (vertical) edges are outgoing (ingoing). Furthermore, 
spectral parameters $\la_i$ and $\mu_k$ are attached to line $i$ and
column $k$. We choose the following parametrization of the
usual Boltzmann weights $a,b$ and $c$,
\begin{eqnarray}
a(\la) &=& 1, \nonumber\\
b(\la) &=& \frac{\sinh( \la-\eta/2)}{\sinh(\la+\eta/2)},
\label{eq:Wpar}\\
c(\la) &=& \frac{\sinh \eta} {\sinh(\la+\eta/2)}.
\nonumber 
\end{eqnarray}
We want to make use of the formalism of the algebraic Bethe
Ansatz. For that purpose the Boltzmann weights are collected in a
matrix $L$,
\begin{equation}
L(\la)=
\left(\begin{array}{@{}cccc@{}}
1&0&0&0\\
0&b(\la)&c(\la)&0\\
0&c(\la)&b(\la)&0\\
0&0&0&1
\end{array}\right).
\end{equation}
The monodromy matrix is then defined as an ordered product of the
$L$-operators,
\begin{equation}
T(\la;\mus) = L(\la-\mu_M)\dots L(\la-\mu_2)L(\la-\mu_1),
\end{equation}
where $L(\la-\mu_k)$ acts on the $k$th factor of the physical space
${\Bbb C}^{2^M}$, and the auxiliary space ${\Bbb C}^{2}$. For clarity,
we will in the following often suppress the explicit dependence of $T$
on $\mus$.  
As an operator on the two-dimensional auxiliary space, $T(\la)$
can be written as
\begin{equation}
T(\la) = 
\left(\begin{array}{@{}cc@{}}
\A(\la) & \B(\la)\\
\C(\la) & \D(\la) 
\end{array}\right)
\label{eq:Taux}
\end{equation}
where $\A$, $\B$, $\C$, $\D$ are operators on the physical space
${\Bbb C}^{2^M}$. The trace of the monodromy operator,
\begin{equation}
\T(\la) = \A(\la) + \D(\la)
\end{equation}
is the usual transfer matrix corresponding to the model with periodic
boundary conditions. Because of the parametrization (\ref{eq:Wpar}),
the monodromy matrix $T(\lambda)$ satisfies the following intertwining
relation,
\begin{equation}
\check{R} (\lambda-\mu) \cdot \left[ T(\lambda) \otimes T(\mu) \right]
= \left[ T(\lambda) \otimes T(\mu) \right] \cdot \check{R}
(\lambda-\mu),
\label{eq:intertwine} 
\end{equation}
The R-matrix $\check{R}$ is defined by,
\begin{equation}
\check{R}(\lambda) = P L(\lambda+\eta/2),
\end{equation}
where $P$ is the permutation operator on the space ${\Bbb C}^{2}
\otimes {\Bbb C}^{2}$. Equation (\ref{eq:intertwine}) embodies
several commutation relations between the operators $\A,\B,\C$ and
$\D$ defined in (\ref{eq:Taux}).

The fixed boundary conditions imply the following
formal expression for the partition function \cite{Kor},
\begin{equation}
Z_M(\las,\mus) = \bra{\downarrow}
\B(\la_1;\mus) \ldots \B(\la_M;\mus) 
\ket{\uparrow},
\end{equation}
where $\ket{\uparrow}$ ($\ket{\downarrow}$) is the state with all
spins up (down). 
In this paper we shall be interested only in the case where $M$ is
even and the $\las$ are chosen as $\{\la_i\}_{i=N+1}^M =
\{\la\}_{i=1}^N$ for $i=1,\ldots,N$, where $N=M/2$. This allows us to
rewrite the partition function in a convenient way \cite{KZJb}. First
define the states,
\begin{equation}
\ket{N} = \B(\la_1)\ldots \B(\la_N)\ket{\uparrow},\quad
\bra{N} = \bra{\uparrow} \C(\la_1)\ldots \C(\la_N).
\end{equation}
Let $\F = \prod_{k=1}^M \sigma_k^x$ be the flip operator on the
physical space that flips all arrows. We then find that,
\begin{eqnarray}
Z_M(\las,\mus) = \bra{N}\F\ket{N}. \label{eq:Partsum}
\end{eqnarray}
Formal expressions for correlation functions can also be written
concisely in this notation. E.g., the probability that all arrows
located at the columns $k_1,\ldots,k_n$ and between the lines $N$ and
$N+1$ are down is given by,
\begin{equation}
\langle \pi_{k_1} \ldots \pi_{k_n} \rangle = 
\frac{\bra{N} \F \pi_{k_1} \ldots \pi_{k_n} \ket{N}} {\bra{N} \F
\ket{N}}, \label{eq:corr}
\end{equation}
where $\pi_k = \frac{1}{2}(1-\sigma_k^z)$. Averages like
(\ref{eq:corr}) can be calculated using the solution of the quantum
inverse scattering problem for the operators $\pi_k$ \cite{MT,GK},
\begin{equation}
\pi_k = \prod_{l=1}^{k-1} \T(\mu_l+\eta/2)\, \D(\mu_k+\eta/2)\,
\prod_{l=k+1}^M \T(\mu_l+\eta/2).
\label{eq:QISMsol}
\end{equation}
From this expression it is clear that the correlation function (\ref{eq:corr}) 
simplifies when the $k_i$ are nearest neighbours. 

\section{Bethe Ansatz}
\label{se:BA}
A marvelous aspect of formul\ae\ (\ref{eq:corr}) and (\ref{eq:QISMsol})
is that the set of inhomogeneities can be chosen such that the state
$\ket{N}$ is an eigenstate of $\T$. In fact, there are many choices
possible that have this property. In this section we will derive
explicit expression for the correlation function (\ref{eq:corr})
corresponding to such choices. We will closely follow a similar
derivation given in \cite{KMT} for one particular choice of the
inhomogeneities, namely those corresponding to the groundstate of the
antiferromagnetic XXZ quantum spin chain. To begin with, we fix the
set of inhomogeneities $\las$ to be a solution of the Bethe Ansatz
equations,
\begin{equation}
\frac{a(\la_j)}{d(\la_j)} \prod_{\scriptstyle
k=1\atop\scriptstyle k\neq j}^N
\frac{b(\la_j-\la_k+\eta/2)}{b(\la_k-\la_j+\eta/2)}  = 1, \qquad 1 \le
j \le N, \label{eq:bethe}
\end{equation}
where $a(\la)=1$ and $d(\la)=\prod_{k=1}^M b(\la- \mu_k)$ 
are the eigenvalues of operators $\A(\la)$ and 
$\D(\la)$ respectively on the reference state $\ket{\uparrow}$. Note
that these equations imply that $\prod_{j=1}^N d(\lambda_j) =1$.
It will be useful to rewrite the Bethe Ansatz equations in their
logarithmic form,
\begin{equation}
\varphi(\lambda_j) = \pi  \pmod{2\pi} ,\quad 1 \leq j \leq N,
\label{eq:BAlog}
\end{equation}
where the function $\varphi$ is defined by,
\begin{eqnarray} 
\varphi(\la) &=& -\i \ln \frac{a(\la)}{d(\la)} - \i \sum_{k=1}^N \ln
\frac{b(\lambda-\lambda_k+\eta/2)} {b(\lambda_k-\lambda
+\eta/2)}  \nonumber\\
&=& \i \ln \frac{d(\la)}{a(\la)} + \i \sum_{k=1}^N \ln  \left( - 
\frac{\sinh(\eta+\lambda-\lambda_k)}{\sinh(\eta-\lambda+\lambda_k)}\right). 
\label{eq:phidef}
\end{eqnarray}
If the $\las$ are a solution of the Bethe Ansatz equations, the state
$\ket{N}$ is a common eigenstate of $\F$ with eigenvalue $\pm 1$, and
of $\T(\la)$ with eigenvalue $t(\la)$ given by,
\begin{eqnarray}
t(\la) &=& a(\la) \prod_{i=1}^N b^{-1} (\la_i-\la+\eta/2) +
d(\la) \prod_{i=1}^N b^{-1} (\la-\la_i+\eta/2) \nonumber\\
&=& a(\la) \left(1 + \e^{ -\i \varphi(\la)} \right) \prod_{i=1}^N
b^{-1}(\la_i-\la+\eta/2). \label{eq:eigvalT}
\end{eqnarray}
If $\las$ obey the Bethe Ansatz equations (\ref{eq:bethe}) and $\{ \xi_j \}$ 
are a set of parameters, then the following holds \cite{Slav,KMT}, 
\begin{equation}
\bra{\uparrow} \prod_{j=1}^N \C(\xi_j) \prod_{i=1}^N \B(\la_j)
\ket{\uparrow} = \frac{\det t'}{\det V}, 
\label{eq:SlavDet}
\end{equation}
where,
\begin{equation}
t'_{ij} = \frac{\der t(\xi_i)} {\der \lambda_j},\quad V_{ij} =
\frac{1}{\sinh(\xi_i -\la_j)}.
\end{equation}
A useful formula that we will use in the following is,
\begin{equation}
\det V \prod_{k,l=1}^N \sinh(\xi_k-\la_l) = \prod_{\scriptstyle
k,l=1\atop\scriptstyle k< l}^N \sinh(\la_k-\la_l) \sinh(\xi_l-\xi_k).
\label{eq:detV}
\end{equation}
From (\ref{eq:Partsum}) it follows that the partition sum is
given by the norm of the Bethe state $\ket{N}$. An expression for the
norm of a Bethe state in terms of a determinant is given by the following
formula which may be obtained by specializing the $\{\xi_j\}$ in
(\ref{eq:SlavDet}) to $\las$, 
\begin{eqnarray}
\bracket{N}{N} &=& \bra{\uparrow} \prod_{i=1}^N
\C(\lambda_i) \prod_{i=1}^N \B(\lambda_i) \ket{\uparrow} \nonumber\\
&=& \sinh(\eta)^N \left( \prod_{\scriptstyle i,j=1\atop\scriptstyle
i\neq j}^N \frac{\sinh(\lambda_i-\lambda_j+\eta)}
{\sinh(\lambda_i-\lambda_j)} \right) \det \varphi',
\end{eqnarray}
where,
\begin{equation}
\varphi'_{ij} = \left. -\i \left( \frac{\der
\varphi(\lambda)}{\der \lambda_j} + \delta_{ij} \frac{\der
\varphi(\lambda)}{\der \lambda} \right) \right|_{\lambda=\lambda_i} 
\label{eq:def_varphip}
\end{equation}
The determinant formula for the norm of a Bethe wavefunction was first
conjectured by Gaudin \cite{G}. Due to the complicated nature of the
Bethe wavefunction a proof was not available till the development of
the quantum inverse scattering method. The first proof of the determinant
formula of the norm of the Bethe wavefunction for the XXZ spin chain was
given by Korepin \cite{Kor}.
  
More work has to be done to obtain an expression for the correlation
function (\ref{eq:corr}). Here we will treat the case when the $k_i$
are nearest neighbours. Using the solution for the quantum inverse
scattering (\ref{eq:QISMsol}) and the fact that $\ket{N}$ is an
eigenstate of $\T(\lambda)$ and $\F$ one finds,
\begin{equation} 
\langle \pi_{k+1}\ldots \pi_{k+n} \rangle = \prod_{j=1}^n
t^{-1}(\mu_{k+j}+\eta/2) \frac{ \bra{N} \prod_{j=1}^n \D(\mu_{k+j}+\eta/2)
\ket{N}} {\bracket{N}{N}}. 
\label{eq:corr2}
\end{equation}
Since $d(\mu_k+\eta/2)=0$, the inverse eigenvalue
$t^{-1}(\mu_{k}+\eta/2)$ takes the simple form,
\begin{equation}
t^{-1}(\mu_{k}+\eta/2) = \prod_{i=1}^N \frac{\sinh(\lambda_i
-\mu_{k}-\eta/2)} {\sinh(\lambda_i -\mu_{k}+\eta/2)}.
\end{equation} 

The action of a product of the operators $\D$ on $\ket{N}$ can be
calculated and is given by,
\begin{equation}
\prod_{j=1}^n \D(\lambda_{N+j}) \prod_{k=1}^N \B(\lambda_k)
\ket{\uparrow} = \sum_{i_1=1}^{N+1}  \sum_{\scriptstyle
i_2=1\atop\scriptstyle i_2\neq i_1}^{N+2} \ldots \sum_{\scriptstyle
i_n=1\atop\scriptstyle i_n\neq i_1,\ldots,i_{n-1}}^{N+n}
G_{i_1,\ldots,i_n} (\las_{i=1}^{N+n}) \prod_{\scriptstyle
k=1\atop\scriptstyle k\neq i_1,\ldots,i_n}^{N+n} \B(\lambda_k)
\ket{\uparrow}, \label{eq:Aprod_onB}
\end{equation}
where the function $G$ is given by,
\begin{equation}
G_{i_1,\ldots,i_n} (\las_{i=1}^{N+n}) = \prod_{l=1}^n d(\lambda_{i_l})
c(\lambda_{i_l}- \lambda_{N+l}+\eta/2) \prod_{\scriptstyle
k=1\atop\scriptstyle k\neq i_1,\ldots,i_l}^{N+l} b^{-1}
(\lambda_{i_l}-\lambda_k+\eta/2).
\end{equation}
We will set $\lambda_{N+j} = \mu_{k+j}+\eta/2$ to calculate the
$n$-point correlation function (\ref{eq:corr2}). Since
$d(\mu_{k})=0$ this means that the sums in (\ref{eq:Aprod_onB}) only
run up to $i_l=N$.

From (\ref{eq:Aprod_onB}) it is seen that we need to calculate the
scalar products of the type,
\begin{equation}
S(\las,\{\la_1,\ldots,\la_{N-n},\mu_{k+1},\ldots,\mu_{k+n}\}) =
\frac{\bra{N} \prod_{i=1}^{N-n} \B(\lambda_i)  \prod_{j=1}^n
\B(\mu_{k+j}+\eta/2) \ket{\uparrow}} {\bracket{N}{N}}.
\end{equation} 
Using (\ref{eq:SlavDet}) and (\ref{eq:detV}) one may express $S$ as a
ratio of determinants, 
\begin{eqnarray}
S(\las,\{\la_1,\ldots,\la_{N-n},\mu_{k+1},\ldots,\mu_{k+n}\}) =
\prod_{\scriptstyle i,j=1\atop\scriptstyle i<j}^n
\frac{\sinh(\la_{N-n+j} -\la_{N-n+i})} {\sinh(\mu_{k+j} -\mu_{k+i})}
\times &&\nonumber\\
\prod_{i=1}^{N-n} \prod_{j=1}^n \frac{\sinh(\la_{i} -\la_{N-n+j})}
{\sinh(\la_{i} -\mu_{k+j}-\eta/2)} 
\prod_{i=1}^{N} \prod_{j=1}^n \frac{\sinh(\la_{i} -\mu_{k+j}+\eta/2)}
{\sinh(\la_{i} -\la_{N-n+j}+\eta)} \frac{\det
\psi'(\las,\{\mu_{k+j}\})} {\det \varphi'(\las)}.&&
\label{eq:Sdetratio}
\end{eqnarray}
The first $N-n$ rows of the $N\times N$ matrix $\psi'$ are equal to
those of the matrix $\varphi'$, but the other $n$ rows are different.
\begin{eqnarray}
\psi'_{ij} &=& \varphi'_{ij},  \quad 1 \leq i \leq N-n,
\label{eq:psidef_f}\\ 
\psi'_{ij} &=&  \frac{\sinh\eta} {\sinh(\lambda_j -\mu_{k+i}-\eta/2)
\sinh(\lambda_j -\mu_{k+i}+\eta/2)}, \quad N-n+1 \leq i \leq
N. \label{eq:psidef_o}   
\end{eqnarray}
Finally, we can rewrite the ratio of the determinants in
(\ref{eq:Sdetratio}) as one determinant by inverting $\varphi'$,
\begin{equation}
\frac{\det \psi'}{\det \varphi'} = \det \left( \psi' \varphi'^{-1}
\right).
\end{equation}

To proceed we have to calculate the matrix $\psi'
\varphi'^{-1}$. The first $N-n$ rows of this matrix can be easily
calculated using (\ref{eq:psidef_f}),
\begin{equation}
\left(\psi' \varphi'^{-1}\right)_{ij} = \delta_{ij},\quad 1\leq i \leq
N-n.
\end{equation}
In the next section we will calculate the other rows in the limit
($M\rightarrow \infty$) for a particular set of solutions of the Bethe
Ansatz equations.


\section{Thermodynamic limit}
\label{se:ThermLim}
\begin{figure}[h]
\centerline{
\begin{picture}(170,100)
\put(0,0){\epsfig{file=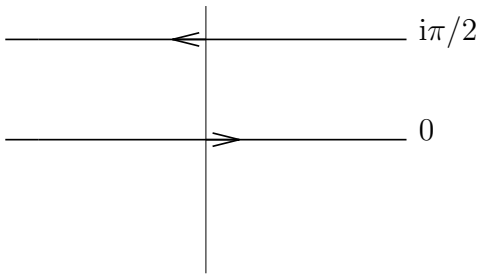}}
\put(155,50){$0$}
\put(155,88){$\i\pi/2$}
\end{picture}}
\caption{The contour ${\mathcal C}$ in the complex plane.}
\label{fig:contour}
\end{figure}

In this section we will describe the thermodynamic limit $M\rightarrow
\infty$. To be able to take this limit we need some information on the
distribution of the solutions of the Bethe Ansatz equations. The
solutions of (\ref{eq:bethe}) fall into two classes depending on the
value of $\eta$. These are the socalled massive regime, where $\Delta
=  \cosh \eta > 1$ and the massless regime where $|\Delta| \leq 1$. In
this paper we will concentrate on the massless case only. Since
$|\Delta| \leq 1$ we will use the parametrization $\gamma = \i\eta$
and furthermore, we will restrict our attention to the interval $\pi/2
> \gamma \geq 0$. We will consider the class of solutions of
(\ref{eq:bethe}) for which the imaginary part of each $\la_j$ is
either $0$ or $\pi/2$. These are the so called 1-strings in the
language of \cite{TS}. In the limit $M\rightarrow \infty$, the
solutions we consider thus belong to a directed contour ${\mathcal C}$ 
(Figure \ref{fig:contour}) which is defined by, 
\begin{equation}
{\mathcal C} = (-\infty,\infty) \cup (\infty + \i\pi/2, -\infty + i\pi/2).
\label{eq:contour}
\end{equation}

Now we will derive the logarithmic form of the Bethe Ansatz
equations in a more precise manner than was done in for
(\ref{eq:BAlog}). For that purpose we define the function $p_n$ by, 
\begin{equation}
p_n (\la) = 
\left\{ \begin{array}{ll}
\dps \hphantom{-}2 \arctan \left(\tanh\la \cot n\gamma/2 \right),
& {\rm for\;} \Im \la=0,\\ 
\dps -2 \arctan \left(\coth\la \tan n\gamma/2 \right), & {\rm for\;} \Im
\la=\pi/2.
\end{array}\right.
\end{equation}
For $\sin n\gamma>0$ this function is monotonously increasing
(decreasing) on the line $\Im \la =0$ $(\Im \la =\pi/2)$. The
logarithmic version of the Bethe Ansatz equations can then be written
as, 
\begin{equation}
\phi(\la_i) = 2\pi n_i, \label{eq:BAlog2}
\end{equation}
where the function $\phi$ is given by,
\begin{equation}
\phi(\la) = \sum_{k=1}^M p_1(\la-\mu_k) - \sum_{j=1}^N
p_2(\la-\la_j).
\end{equation}
The numbers $n_i$ appearing in the right hand side of (\ref{eq:BAlog2}) are
integers for $N$ odd and half integers for $N$ even. For every (half) integer
$\{n_i\}$ there are two solutions of the Bethe Ansatz equations,
corresponding to the two different values of the imaginary part. A
solution thus is uniquely specified by a set of integers and a
corresponding set of parities, where the parity of a solution is defined by,
\begin{equation}
v = 1- \frac{4}{\pi} \Im \la.
\end{equation}
Given a set of (half) integers $\{n_i\}$ and a
set of parities $\{v_i\}$, a solution $\la_j$ of (\ref{eq:BAlog2}) is
called a particle. A solution $\la_{\rm h} = \bla_{\rm h} +
\i\pi(1-v_{\rm h})/4$ to the equation,   
\begin{equation}
\phi(\la_{\rm h}) = 2\pi m,\qquad m \not\in \{n_i\}\;
{\rm or}\; v_{\rm h} \not\in \{v_i\},
\end{equation}
is called a hole. In the thermodynamic limit, the particles and holes
have finite distribution densities $\rho_{\rm p}$ and $\rho_{\rm h}$,
defined by,  
\begin{eqnarray}
M \rho_{\rm p}(\la) \d\la &\quad {\rm number\; of\;
particles\; in}& \quad [\la,\la+\d\la], \\
M \rho_{\rm h}(\la) \d\la &\quad {\rm number\; holes\; in}&
\quad [\la,\la+\d\la]. 
\end{eqnarray}
Note that the 1-form $\d \la$ has a direction corresponding to 
that of ${\mathcal C}$. The density $\rho_{\rm tot}$ of the total 
possible solutions, or vacancies, is given simply by,
\begin{equation}
\rho_{\rm tot}(\la) = \rho_{\rm p}(\la) + \rho_{\rm h}(\la).
\end{equation}
Since we are dealing with an inhomogeneous model, it will be useful to
define the densities $\tilde{\rho}_{\rm tot}$ by,
\begin{equation}
\rho_{\rm tot}(\lambda) = \frac{1}{M} \sum_{k=1}^M \tilde{\rho}_{\rm
tot} (\lambda-\mu_k), \label{eq:localdens} 
\end{equation}
The corresponding particle and hole densities are given by,
\begin{eqnarray}
\tilde{\rho}_{\rm p} (\la-\mu_k) &=& \vartheta(\la)
\tilde{\rho}_{\rm tot}(\la-\mu_k), \label{eq:rhoptdef} \\
\tilde{\rho}_{\rm h} (\la-\mu_k) &=& (1-\vartheta(\la))
\tilde{\rho}_{\rm tot}(\la-\mu_k),
\end{eqnarray}
where the Fermi weight $\vartheta(\la)$ is given by,
\begin{equation}
\vartheta(\la) = \frac{\rho_{\rm p}(\la)}{\rho_{\rm tot}(\la)}.
\label{eq:Fermiwdef}
\end{equation}
Using these densities we can take the limit $M\rightarrow \infty$. It
follows that (\ref{eq:BAlog2}) in the thermodynamic limit can be
written as,
%
\begin{equation}
\lim_{M\rightarrow \infty} \frac{1}{M} \phi(\la) = 
\pi \left( -1 + 2 \int_{-\infty}^{\la} \rho_{\rm tot}(\la')
\d\la'\right),
\label{eq:BAlog_c}
\end{equation}
where the integration is along the contour ${\mathcal C}$, Figure 
\ref{fig:contour}. Differentiating with respect to $\la$ 
we find the formula, 
\begin{equation}
\rho_{\rm tot}(\la) = K_1{}^{\rm tot}(\la)  -
\int_{\mathcal C} K_2(\la-\la') \rho_{\rm p}(\la') \d\la',
\label{eq:BAinfc} 
\end{equation}
The function $K_1^{\rm tot}$ is defined by,
\begin{equation}
K_1{}^{\rm tot}(\la) = \lim_{M \rightarrow \infty} \frac{1}{M}
\sum_{k=1}^M K_1(\la-\mu_k), 
\end{equation}
and $K_n$ is given by,
\begin{equation}
K_n(\la) = \frac{1}{2\pi} p'_n(\la) = \frac{1}{2\pi} \frac{\sin n
\gamma} {\sinh(\la-\i n\gamma/2)\sinh(\la+\i n
\gamma/2)}. \label{eq:Kdef}
\end{equation}
The thermodynamic limit of $\varphi'$ is found using
(\ref{eq:BAlog_c}) and the fact that $\phi(\la) = \varphi(\la) \bmod{\pi}$, 
\begin{equation}
\varphi_{ij}' = -2\pi \i \left(M \delta_{ij} \rho_{\rm
tot}(\la_i) + K_2(\la_i -\la_j) \right).
\label{eq:philim_c}
\end{equation}

Now we are in a postion to return to the calculation at the end of the
previous section. Remember that we want to calculate the last $n$ rows of
the matrix  $\psi'\varphi'^{-1}$. For that purpose we recall from
(\ref{eq:psidef_o}) that,
\begin{eqnarray}
\psi'_{N-n+i,j} &=& -2 \pi\i K_1(\la_j-\mu_{k+i})
\nonumber\\
&=& -2\pi\i \left( \tilde{\rho}_{\rm tot}(\la_j-\mu_{k+i})
+ \int_{\mathcal C} K_2(\la_j-\la') \vartheta(\la') \tilde{\rho}_{\rm
tot}(\la'-\mu_{k+i}) \d\la' \right), 
\end{eqnarray}
where in the second line we have used (\ref{eq:rhoptdef}) and
(\ref{eq:BAinfc}). From (\ref{eq:philim_c}) however and the fact that
$K_2$ is symmetric it follows that this is precisely equal to,
\begin{equation}
\psi'_{N-n+i,j} = \frac{1}{M} \sum_{l=1}^N \frac{\tilde{\rho}_{\rm tot}
(\la_l -\mu_{k+i})} {\rho_{\rm tot}(\la_l)}
\varphi'_{lj}.  
\end{equation}
We thus conclude that,
\begin{eqnarray}
\left( \psi' \varphi'^{-1} \right)_{ij} &=& \delta_{ij},\quad 1 \leq i
\leq N-n \\
\left( \psi' \varphi'^{-1} \right)_{N-n+i,j} &=&
\frac{\tilde{\rho}_{\rm tot} (\la_j -\mu_{k+i})} {M \rho_{\rm
tot}(\la_j)}, \quad 1 \leq i \leq n.
\end{eqnarray}
The determinant of this matrix can be written concisely as,
\begin{equation}
\det \left(\psi' \varphi'^{-1} \right) = \det
\tilde{S} \frac{1}{M^n} \prod_{j=1}^n \rho_{\rm
tot}^{-1}(\la_{N-n+j}), \label{eq:detpsiphi} 
\end{equation}
where the $n\times n$ matrix $\tilde{S}$ is given by,
\begin{equation}
\tilde{S}_{ij} = \tilde{\rho}_{\rm tot}(\la_{N-n+j} -
\mu_{k+i}). \label{eq:Sdef}
\end{equation}

Finally, using (\ref{eq:corr2}), (\ref{eq:Aprod_onB}),
(\ref{eq:Sdetratio}) and (\ref{eq:detpsiphi}), the emptiness formation
probability can be written as,  
\begin{equation}
\langle \pi_{k+1} \dots \pi_{k+n} \rangle = \frac{1}{\displaystyle
M^n \prod_{l<m} \sinh(\mu_{k+l} -\mu_{k+m})} \sum_{i_1=1}^N \dots
\sum_{i_n=1}^N H( \{\la_{i_l}\}, \{\mu_{k+l}\}) \prod_{l=1}^n \rho_{\rm
tot}^{-1}(\la_{i_l}), \label{eq:efp1}
\end{equation}
where the function $H$ is given by,
\begin{eqnarray}
H( \{\la_{i_l}\}, \{\mu_{k+l}\}) &=& \frac{\det
\tilde{S}(\{\la_{i_l}\}, \{\mu_{k+l}\})}{\dps \prod_{l<m}
\sinh(\la_{i_m} -\la_{i_l}-\i\gamma)} 
\times\nonumber\\
&&\hspace{-2.5cm}\prod_{l=1}^n \left(\prod_{m=1}^{l-1}
\sinh(\la_{i_l}-\mu_{k+m}-\i\gamma/2) \prod_{m=l+1}^{n}
\sinh(\la_{i_l}-\mu_{k+m}+\i\gamma/2) \right).
\end{eqnarray}
The last step in deriving an
expression for the emptiness formation probability in the
thermodynamic limit is to replace the sums in (\ref{eq:efp1}) by
integrals using the discussion above. We then arrive at the following
multiple integral expression, 
\begin{equation}
\langle \pi_{k_1} \dots \pi_{k_n} \rangle = \frac{1}{\displaystyle
\prod_{l<m} \sinh(\mu_{k+l} -\mu_{k+m})} \int_{\mathcal C} \ldots
\int_{\mathcal C} H( \{\la_l\},\{\mu_{k+l}\}) \prod_{l=1}^n
\vartheta(\la_l) \d\la_l,   
\end{equation}
where the Fermi weight $\vartheta(\la)$ is define in
(\ref{eq:Fermiwdef}). We remind the reader that the integration
is along the directed contour ${\mathcal C}$, equation (\ref{eq:contour}),
Figure \ref{fig:contour}.

\section{Conclusion}
In this paper we have obtained a multiple integral expression for the emptiness 
formation probability (EPF) on the central horizontal line of the inhomogeneous 
six-vertex model with domain wall boundaries. We derived this expression in the 
thermodynamic limit when the inhomogeneities are chosen from a particular set of 
solutions of the Bethe Ansatz equations, namely those without a bound states but 
otherwise arbitrary. This result is a first step to obtain an expression for the 
EPF for general solutions of the Bethe Ansatz equations, i.e. also for bound state 
solutions. We expect that certain properties of the EPF are independent of the 
special choice of inhomogeneities. Ultimately we hope to learn more about such 
properties by studying the EPF averaged over Bethe Ansatz solutions.

\section{Acknowledgements}
We thank R. Martinez for his involvement in this paper. This work has been supported 
by the Australian Research Council (ARC) and by the National Science Foundation 
under grant number PHY-9605226.

%

\end{document}